\pdfoutput=1%
\documentclass[a4paper,american,floatfix,pdftex,superscriptaddress,twoside,%
aps,prl,%
citeautoscript,%
longbibliography,%
reprint,twocolumn,%,final% add final to see real layout, no todonotes anymore, ...
]{revtex4-2}%

\PassOptionsToPackage{unicode}{hyperref} % please leave linkframes in the file for online viewing, arXiv, etc. -- they help the reader
\usepackage{amsfonts,amsmath,amssymb}
\usepackage{graphicx}
\usepackage[T1]{fontenc}
\usepackage[utf8]{inputenc}
\usepackage{hypernat}
\usepackage[ams,todos]{CMImacros}

\newcommand*\IB{\ensuremath{\text{IB}}\xspace} % iodobenzene
\newcommand*\DFIB{\ensuremath{\text{DFIB}}\xspace} % difluoroiodobenzene

\graphicspath{{./fig/}{./}}
\newcommand{\cfeldesy}{\affiliation{Center for Free-Electron Laser Science CFEL, Deutsches
    Elektronen-Synchrotron DESY, Notkestr. 85, 22607 Hamburg, Germany}}
\newcommand{\uhhcui}{\affiliation{Center for Ultrafast Imaging, Universität Hamburg, Luruper
      Chaussee 149, 22761 Hamburg, Germany}}%
\newcommand{\uhhphys}{\affiliation{Department of Physics, Universität Hamburg, Luruper Chaussee 149,
    22761 Hamburg, Germany}}%\
\newcommand{\uhhchem}{\affiliation{Department of Chemistry, Universität Hamburg,
    Martin-Luther-King-Platz 6, 20146 Hamburg, Germany}}%
\newcommand{\stemail}{\email[Email:~]{sebastian.trippel@cfel.de}}%
\newcommand{\cmiweb}{\homepage[\\Website:~]{https://www.controlled-molecule-imaging.org}}%

\newcommand*{\varxT}{\ensuremath{\frac{1}{4}\sqrt{\frac{\pi}{\Delta\omega_{\chi,\Trot}}}}\xspace}
\newcommand*{\varzT}{\ensuremath{\frac{1}{4}\sqrt{\frac{\pi}{\Delta\omega_{\upsilon,\Trot}}}}\xspace}

\newcommand*{\elfieldvec}{\ensuremath{\boldsymbol{E}}\xspace}
\newcommand*{\elfieldnought}{\ensuremath{E_0}\xspace}
\newcommand*{\xvec}{\ensuremath{\boldsymbol{x}}\xspace}
\newcommand*{\yvec}{\ensuremath{\boldsymbol{y}}\xspace}

\renewcommand*{\Ialign}{\ensuremath{I}\xspace}

\begin{document}
\title{Analytic-harmonic-librator representation of adiabatic 3D alignment \\ of asymmetric-top
   molecules}%
\author{Sebastian~Trippel}\stemail\cmiweb\cfeldesy\uhhcui%
\author{Nidin~Vadassery}\cfeldesy\uhhchem%
\author{Jochen~Küpper}\cfeldesy\uhhcui\uhhchem\uhhphys%
\begin{abstract}\noindent%
   Achieving 3D alignment of gas-phase molecules enables direct molecular-frame measurements by
   bypassing rotational averaging. Here, we analytically describe the adiabatic alignment during the
   laser pulse of asymmetric-top molecules in elliptically polarized fields within the
   harmonic-librator limit. The derived expressions characterize the angular probability
   distributions for all three Cardan angles, applicable to thermal ensembles at low and high
   temperatures. Validated by numerical results, these compact equations simplify the interpretation
   of 3D alignment experiments and elucidate its dependence on molecular parameters such as
   rotational constants and polarizability anisotropies as well as laser properties and temperature.
\end{abstract}
\date{\today}%
\maketitle%

%Introduction
Controlling the rotational degrees of freedom of molecules through angular confinement serves a
broad spectrum of purposes in the molecular sciences. Key applications include the investigation of
steric effects in chemical reactivity~\cite{Brooks:Science193:11} as well as the detailed probing of
dynamics in the molecular frame~\cite{Kelkensberg:PRA84:051404, Bisgaard:Science323:1464,
  Holmegaard:NatPhys6:428, Kraus:PRL111:243005, Boll:FD171:57, Trabattoni:NatComm11:2546}. Gaining
insight into molecular frame properties is critical for various modern approaches used to observe
the real-time dynamics of nuclei and electrons in chemical processes~\cite{Kuepper:PRL112:083002,
  Barty:ARPC64:415, Hensley:PRL109:133202}. Techniques like x-ray~\cite{Kuepper:PRL112:083002,
  Kierspel:JCP152:084307, Pabst:PRA81:043425, Debnarova:JCP134:054302} and
electron~\cite{Meckel:Science320:1478, Hensley:PRL109:133202, Pullen:NatComm6:7262,
  Yang:NatComm6:8172, Walt:NatComm8:15651} diffraction enable imaging molecular structure and
dynamics directly in the molecular frame, when combined with alignment or orientation. Furthermore,
laser alignment was successfully implemented in high-harmonic generation
spectroscopy~\cite{Torres:PRL98:203007, Weber:PRL111:263601} and strong-field
ionization~\cite{Pavicic:PRL98:243001, Meckel:NatPhys10:594}. Alignment and orientation also provide
a pathway to resolve the structure of molecular orbitals~\cite{Itatani:Nature432:867,
  Holmegaard:NatPhys6:428, Vozzi:NatPhys7:822}. Such imaging techniques are essential for studying
complex systems, particularly asymmetric-top molecules, which constitute the vast majority of
molecular species. While our treatment assumes parallel principal axes of inertia and
polarizability, the results remain approximately valid to first order when these axes are nearly
aligned, as demonstrated elsewhere for similar
systems~\cite{Hansen:JCP139:234313,Mullins:NatComm13:1431}.

Laser-induced alignment of gas-phase molecules is a well-established method for accessing the
molecular frame~\cite{Stapelfeldt:RMP75:543}. However, the persistent laser field can interfere with
the dynamics under study~\cite{Meyer:PRL108:063007}. This prompted the development of techniques for
achieving alignment and orientation in field-free space~\cite{Stapelfeldt:RMP75:543,
   RoscaPruna:PRL87:153902}. While these methods work effectively for linear
molecules~\cite{Ghafur:NatPhys5:289, Karamatskos:NatComm10:3364, Trippel:PRL114:103003}, achieving
strong, especially 3D, field-free alignment for asymmetric top molecules remains a
challenge~\cite{Larsen:PRL85:2470, Underwood:PRL94:143002, Lee:PRL97:173001,
   Holmegaard:PRA75:051403R, Ren:PRL112:173602}. One solution is to use adiabatic alignment followed
by a rapid switch-off of the laser field~\cite{Yan:JCP111:4113, Seideman:JCP115:5965,
   Underwood:PRL90:223001, Sussman:PRA73:053403, Goban:PRL101:013001, Chatterley:JCP148:221105,
   Mullins:NatComm13:1431}. This provides a picosecond time window of field-free alignment before
the rotational wave packet propagates in such an extent that the alignment decays.

Analytical solutions of the properties of molecules in strong electric fields date back to initial
investigations of linear molecules in static electric fields~\cite{Peter:JCP26:1657,
   Bulthuis:JCSFT91:205}. These investigations were later extended first to linear
molecules~\cite{Friedrich:PRL74:4623, Seideman:JCP115:5965} and then to symmetric
tops~\cite{Kim:JCP108:6763, Haertelt:JCP128:224313} in linearly polarized laser fields. Here,
analytical expressions were developed, in particular, for small molecules at low temperatures and
large molecules at high temperatures~\cite{Seideman:JCP115:5965, Kumarappan:JCP125:194309,
   Friedrich:EPJD38:209}. In addition, analytical expressions for asymmetric top molecules in DC
fields were derived~\cite{Kanya:PRA70:013403}.

3D alignment can be provided, among other means, by elliptically polarized light. While numerical
quantum-mechanical simulations are available for this case, the complexity of the approach precludes
a simple methodological access to the relevant parameters. Such a model is essential for a
predictive understanding of three-dimensional (3D) molecular alignment. Furthermore, numerical
simulations of large asymmetric top molecules are computationally challenging, even at room
temperature. Analytical descriptions with well-defined asymptotic limits facilitate the description
of the classical limit of such systems~\cite{Zhao:PRA109:053107, Amin:JACS147:7445}.

The development of increasingly powerful lasers~\cite{Deppe:OptExp23:28491} will open up new
possibilities for 3D alignment of both small molecules and large systems such as proteins, viruses,
or nanocrystals~\cite{Amin:JACS147:7445, Sinelnikova:BPJ120:3709, Agelii:PCCP27:10939}. For example,
x-ray diffraction experiments on laser-oriented crystals were recently
conducted~\cite{Gao:PNAS116:4018}. The use of elliptically polarized light for 3D alignment of large
systems could provide ideal samples and has direct implications for sample delivery in XFEL-based
single-particle imaging (SPI)~\cite{Neutze:Nature406:752}. % I would amend the citation by Lukas' manuscript once public (arXiv)
A simple and intuitive access to the angular distributions is thus highly desirable.

\begin{figure}
   \includegraphics[width=\linewidth]{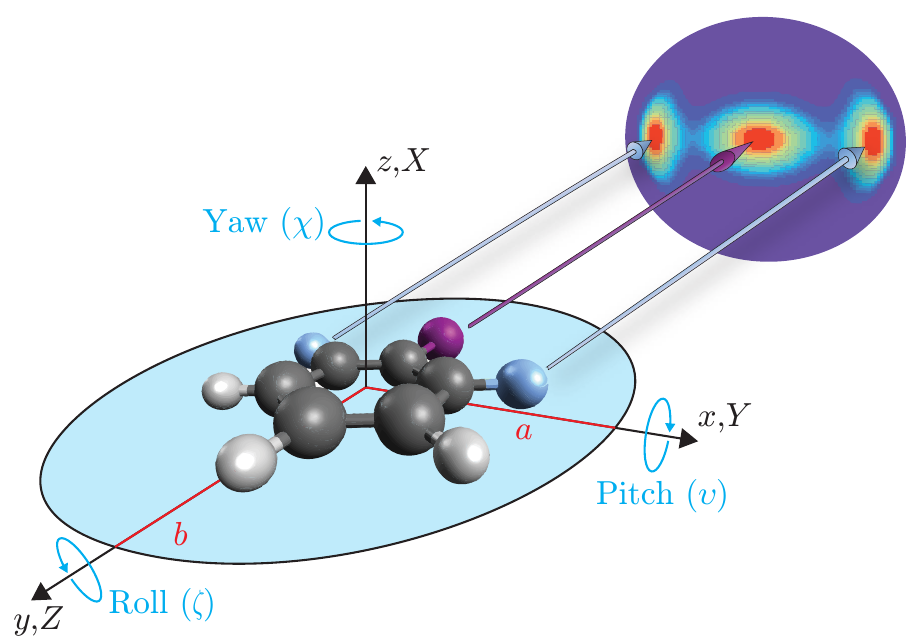}
   \caption{3D-aligned \DFIB, projection onto a detector for I and F atoms and definition of the
      laboratory coordinate system with rotation angles. The time dependent elliptically polarized
      electric field $\elfieldvec(t)$ with coefficients $a$ and $b$ is rotating in the $x$--$y$
      plane. The alignment-light field is propagating along the laboratory $z$ axis.}
   \label{fig:coordinate-system}
\end{figure}

% Theory
We consider an asymmetric top molecule with a polarizability tensor that is diagonal in the
principal-axes-of-inertia system exposed to an elliptically polarized light field, as shown in
\autoref{fig:coordinate-system} for the prototypical 2,6-difluoro-iodobenzene (\DFIB) as an example.
The laser is assumed to propagate along the $z$ direction and its instantaneous electric field
vector $\elfieldvec(t)$ sweeps over an area represented by the light blue ellipse as a function of
time $t$. In our case, the electric field vector is given by
\begin{equation}
  \elfieldvec(t) = \elfieldnought(t)\left(a\hat{\xvec}\cos\omega t + b\hat{\yvec}\sin\omega t \right).
\end{equation}
The angular frequency is denoted as $\omega$, $a^2+b^2=1$, and the time-dependent electric field
amplitude in $x$ and $y$ direction is given by $a\elfieldnought(t)$ and $b\elfieldnought(t)$. Unit
vectors along $x$ and $y$ directions are denoted as $\hat{\xvec}$ and $\hat{\yvec}$,
respectively. In addition, we assume all time scales of the laser-pulse envelope to be large enough
to provide adiabatic alignment. The molecular rotational time scale is furthermore assumed to be
much longer than one optical half-cycle of the laser pulse. Therefore, the interaction can be
averaged over one optical period, and $\elfieldnought(t)$ is replaced with
$\elfieldnought/\sqrt{2}$. The coordinate system in the molecular frame is chosen such that for the
principal moments of polarizability $\alpha_{ZZ}>\alpha_{YY}>\alpha_{XX}$. In this scenario, the
molecule 3D aligns as depicted in \autoref{fig:coordinate-system} when $b>a$.

To describe the 3D alignment of molecules in the harmonic-librator limit, the corresponding
field-dependent Hamiltonian is expanded into a Taylor series in angle coordinates around the
potential minimum. The interaction potential in the coordinate system defined above is given
by~\cite{Rouzee:PRA77:043412}:
\begin{eqnarray}
  V_\mathrm{int}&=&\frac{1}{4}\elfieldnought^2\biggl\{\Delta\alpha_{ZY}\left[a^2\cos^2\theta_{zZ}-(b^2-a^2)\cos^2\theta_{yZ}\right]\nonumber\\
  &&- \Delta\alpha_{YX}\left[a^2\cos^2\theta_{zX}-(b^2-a^2)\cos^2\theta_{yX}\right]\biggr\},
  \label{eqn:V_int}
\end{eqnarray}
where $\Delta\alpha_{\Gamma\Gamma'} = \alpha_{\Gamma\Gamma}-\alpha_{\Gamma'\Gamma'}$ and
$\cos\theta_{\gamma\Gamma}$ are the direction cosines with $\gamma = x, y, z$, $\Gamma,\Gamma' = X, Y, Z$.
Lowercase subscripts denote variables in the laboratory frame, while uppercase subscripts refer to
the molecular frame~\footnote{Following Stapelfeldt and Rouzée~\cite{Stapelfeldt:RMP75:543,
      Rouzee:PRA77:043412}, we use uppercase labels for the molecular frame and lowercase labels for
   the laboratory frame, reversing the usual spectroscopic convention. Our molecular-axis assignment
   $(X,Y,Z)=(c,b,a)$ corresponds to representation $\mathrm{I}^{l}$ in the notation of Gordy and
   Cook (Table~7.3).} The direction cosines are transferred to Cardan angles $\chi$, $\upsilon$, and
$\zeta$, defined with respect to the perfectly aligned molecule along the laboratory axis, as
displayed in \autoref{fig:coordinate-system}. We use Cardan angles (instead of Euler angles) because
they avoid singularities for asymmetric tops in elliptically polarized fields, and the Hamiltonian
becomes diagonal in this basis, enabling separation of the $\chi$, $\upsilon$, and $\zeta$ motions
in the harmonic-librator limit. To obtain the curvature of the potential in the harmonic-librator
limit \eqref{eqn:V_int} is expanded, for small angles, in Cardan angles. Up to an angle-independent
energy offset this results in

\begin{eqnarray}
  V_\mathrm{int} &\approx& -\frac{1}{4}\elfieldnought^2\biggl\{\Delta\alpha_{ZX}b^2(1-\upsilon^2) \\
                 &&+ \Delta\alpha_{YX}a^2(1-\zeta^2) + \Delta\alpha_{ZY}(b^2-a^2)(1-\chi^2)
                    \biggr\}.\nonumber
  \label{eqn:V_int_harm_osc}
\end{eqnarray}
For small angles, the interaction is a three-dimensional anisotropic harmonic oscillator potential
in the coordinates $\chi$, $\upsilon$, and $\zeta$. Due to the assumed molecular symmetry,
cross-terms vanish. In the harmonic-librator limit, the reduced dimensionless Hamiltonians for the
$\zeta$, $\upsilon$, and $\chi$ motions are obtained by scaling each coordinate-specific Hamiltonian
by their corresponding rotational constant $A$, $B$, or $C$:
\begin{gather}
  \label{eq:hamiltons}
  \hat{H}_{\xi} = -\frac{\partial^2}{\partial \xi^2}-\Delta\omega_\xi (1-\xi^2) \text{ with } \xi = \chi, \upsilon, \zeta;\\
  \label{eq:omega_zeta}
  \Delta\omega_\zeta \equiv \frac{\Delta\alpha_{YX}\elfieldnought^2}{4A}a^2\\
  \label{eq:omega_upsilon}
  \Delta\omega_\upsilon \equiv \frac{\Delta\alpha_{ZX}\elfieldnought^2}{4B}b^2\\
  \label{eq:omega_chi}
  \Delta\omega_\chi \equiv \frac{\Delta\alpha_{ZY}\elfieldnought^2}{4C}(b^2-a^2)
\end{gather}
The three reduced interaction parameters are defined by \eqref{eq:omega_zeta}--\eqref{eq:omega_chi}
as $\Delta\omega_\zeta$, $\Delta\omega_\upsilon$, and $\Delta\omega_\chi$, respectively. They depend
on the three possible polarizability volume differences as well as on the field strength along the
respective molecular axis, as encoded in the ellipticity of the light field.

The functional form of each reduced Hamiltonian corresponds to a reduced harmonic oscillator
Hamiltonian with an angle-independent offset. Its wave function in all three angular dimensions has
the form $\psi_{\nu_\xi} = NH_{\nu_\xi}(u)\exp{-u^2/2}$. $N$ denotes a normalization factor,
$H_{\nu_\xi}(u)$ are the Hermite polynomials, $\nu_\xi=0,1,2\ldots$, and
$u=\Delta\omega_\xi^{1/4}\xi$ with the corresponding reduced interaction parameter
$\Delta\omega_\xi$. The pendular limit expressions for the eigenenergies, excluding the global
angle-independent energy shift $-C\Delta\omega_\chi - B\Delta\omega_\upsilon - A\Delta\omega_\zeta$,
are given by those of the anisotropic 3D quantum harmonic oscillator:
\begin{eqnarray}
  E_{\nu_\chi,\nu_\upsilon,\nu_\zeta} &=& 2\sqrt{\Delta\omega_\zeta} A \left(\nu_\zeta + \frac{1}{2}\right)\nonumber\\
  &+& 2\sqrt{\Delta\omega_\upsilon} B \left(\nu_\upsilon + \frac{1}{2}\right)\\
  &+& 2\sqrt{\Delta\omega_\chi} C \left(\nu_\chi + \frac{1}{2}\right),\nonumber
\end{eqnarray}
with the harmonic oscillator quantum numbers $\nu_\chi, \nu_\upsilon, \nu_\zeta = 0, 1, 2 \ldots$.

The state specific angular variances are given by
$\expectation{\xi^2}=(\nu_\xi+1/2)/\sqrt{\Delta\omega_\xi}$. This yields
\begin{equation}
  \langle \cos^2\xi \rangle \approx 1-\langle \xi^2 \rangle = 1-
  \frac{\nu_\xi+\frac{1}{2}}{\sqrt{\Delta\omega_\xi}}.
\end{equation}
For all angles, the variances of the angular distributions decrease with increasing reduced
interaction parameter, that is, the molecule gets more strongly confined, and the corresponding
degree of alignment converges toward 1. For linearly polarized light ($b=1$, $a=0$), the azimuthal
angle $\zeta$ (rotation about the $y$-axis) is uniformly distributed, implying the absence of
confinement about this axis and isotropic orientation in the $xz$-plane, \ie,
$\langle\cos\zeta\rangle=0$ and $\langle\cos^2\zeta\rangle=1/2$.

To connect with literature using Euler angles, we transform the expectation values of Cardan-angle
functions. For small angles, we derive \textcosthreeD, defined as the angle between the C--I axis of
\DFIB{} and the laboratory $y$-axis:
\begin{equation}
  \textcosthreeD \approx 1 - \frac{\nu_\upsilon + \frac{1}{2}}{\sqrt{\Delta \omega_\upsilon}} -
  \frac{\nu_\chi + \frac{1}{2}}{\sqrt{\Delta \omega_\chi}},
  \label{eqn:cost-ellipt-states}
\end{equation}
with $\nu_\upsilon, \nu_\chi = 0, 1, 2, \ldots$. For linearly polarized light and
prolate-symmetric-top molecules ($C = B$), the interaction parameters satisfy
$\Delta \omega_\upsilon = \Delta \omega_\chi$. In this limit, the two terms in
\eqref{eqn:cost-ellipt-states} merge, and the expression reduces to:
\begin{equation}
  \textcosthreeD \rightarrow 1 - \frac{J + 1}{\sqrt{\Delta \omega_\upsilon}},
  \label{eqn:cost_lin}
\end{equation}
where $J = \nu_\upsilon + \nu_\chi$ is the total angular momentum quantum number; the sum of the
zeropoint energies accounts for the 2D isotropic harmonic oscillators' contributions. This is the
well-known expression for symmetric-top or linear molecules with
$J = 0, 1, 2, \ldots$~\cite{Seideman:JCP115:5965,Friedrich:PRL74:4623}.

For larger molecules it is desirable to use the initial field-free rotational temperature $\Trot$
instead of the rotational constants and quantum numbers~\cite{Seideman:JCP115:5965,
  Friedrich:EPJD38:209, Kumarappan:JCP125:194309}. For linear molecules with rotational constant
$B$, the results are expected to describe the degree of alignment well for $\kB \Trot/B \gg
1$~\cite{Seideman:JCP115:5965}. Rotational temperature-dependent reduced interaction parameters can
be defined as $\Delta\omega_{\chi,\Trot} = \Delta\alpha_{ZY}\elfieldnought^2(b^2-a^2)/4\kB \Trot$,
$\Delta\omega_{\upsilon,\Trot} = \Delta\alpha_{ZX}\elfieldnought^2b^2/4\kB \Trot$, and
$\Delta\omega_{\zeta,\Trot} = \Delta\alpha_{YX}\elfieldnought^2a^2/4\kB \Trot$, respectively. For
the large-molecule high-temperature case, one obtains in the harmonic-librator limit for asymmetric
top molecules in elliptically polarized light fields
\begin{equation}
  \costhreeDaveraged = 1 - \varzT - \varxT.
  \label{eqn:approx_temp_first_order_ellipt}
\end{equation}
For symmetric top molecules in linearly polarized light, it reduces to
\begin{equation}
  \costhreeDaveraged \rightarrow
  1-\frac{1}{2}\sqrt{\frac{\pi}{\Delta\omega_{\upsilon,\Trot}}},
  \label{eqn:approx_temp_first_order_lin}
\end{equation}
matching the expressions derived in~\cite{Seideman:JCP115:5965, Friedrich:EPJD38:209}. The alignment
is determined by the competition between the coupling strength and the rotational temperature, as
already established for linear molecules in linear light fields~\cite{Seideman:JCP115:5965}.
Structural characteristics are effectively reduced to polarizability anisotropies, eliminating any
dependence on principal moments of inertia. Conversely, in the low-temperature regime for light
molecules, the degree of alignment is governed by a competition between the interaction strength and
the rotational constant.

\begin{figure}
   \includegraphics[width=\linewidth]{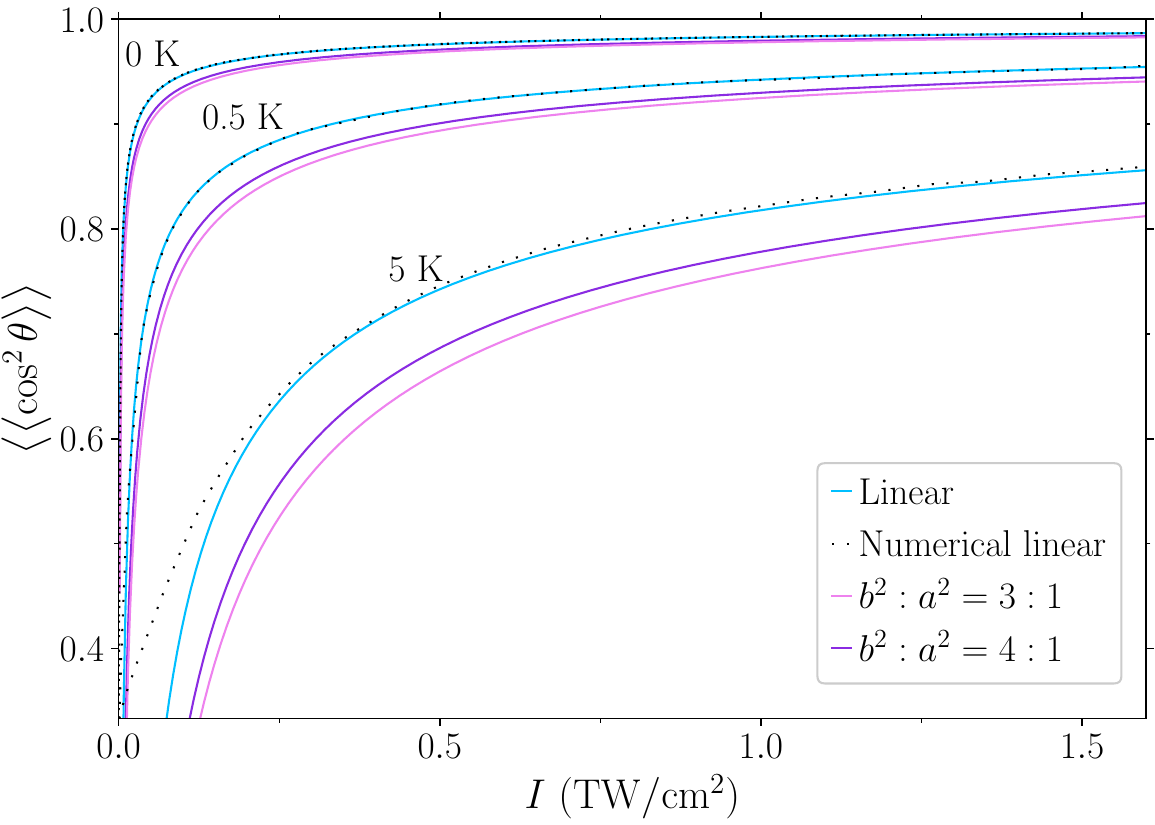}%
   \caption{Alignment of \IB for various temperatures and ellipticities as a function of the
      cycle-averaged laser intensity $I$. Numerical results are shown as dotted lines.}
   \label{fig:IB-alignment-linear}
\end{figure}

% Results and discussion
To test the validity of the derived analytical equations, we compared with numerical results for
iodobenzene (\IB) and \DFIB~\footnote{ For \IB: $\alpha_{XX} = 10.1$~\AA$^3$,
   $\alpha_{YY} = 15.0$~\AA$^3$, $\alpha_{ZZ} = 20.9$~\AA$^3$; $A = 5669.126$~MHz,
   $B = 750.414323$~MHz,
   $C = 662.636162$~MHz~\cite{Dorosh:JMolSpec246:228,Kumarappan:JCP125:194309}. For \DFIB:
   $\alpha_{XX} = 8.5$~\AA$^3$, $\alpha_{YY} = 14.5$~\AA$^3$, $\alpha_{ZZ} = 21.3$~\AA$^3$;
   $A = 1740$~MHz, $B = 713$~MHz, $C = 506$~MHz~\cite{Nevo:PCCP11:9912}. For OCS:
   $\alpha_\perp = 3.74$~\AA$^3$, $\alpha_\parallel = 7.79$~\AA$^3$;
   $B=6081.492$~MHz~\cite{Elliott:PRL46:317,Gussoni:JMolStruct447:163,Maki:JPCRD3:221}. }.
\autoref{fig:IB-alignment-linear} shows the degree of alignment for \IB for various temperatures and
ellipticities.
Here, \eqref{eqn:cost-ellipt-states} and \eqref{eqn:approx_temp_first_order_ellipt} were used for
$\Trot=0$~K and $\Trot\ne 0$~K, respectively. The corresponding numerical results for $\Trot=0$~K
were obtained using TDasyrot, RichMol, or LIMAO~\cite{Omiste:PRA94:063408, Owens:JCP148:124102,
   Szidarovszky:ComPhysComm228:219} for linearly polarized light, reproducing the analytical
solution. The numerical results for $\Trot \ne 0$ were reproduced
from~\cite{Kumarappan:JCP125:194309}. For all temperatures, the typical increase in the degree of
alignment with increasing laser intensity is observed. The degree of alignment decreases with
increasing $\Trot$. Furthermore, $\costhreeDaveraged$ is reduced when changing from linearly
polarized light to elliptical polarization as that reduces the field strength along the laboratory
$y$ axis and thus the corresponding confinement of the molecular
$Z$-axis~\cite{Viftrup:PRL99:143602}. The linear case exhibits an excellent agreement between
numerical results and the harmonic-librator limit for $\costhreeDaveraged\gtrsim0.6$ for all
temperatures. The slight deviation at 5~K in the pendular limit could potentially be due to an
insufficient number of rotational states included in the numerical
calculation~\cite{Kumarappan:JCP125:194309}, which leads to a slightly increased numerical
$\costhreeDaveraged$.

The accuracy of the analytical expressions stems from the robustness of the degree of alignment
across different initial rotational states. Although pendular eigenenergies, \eg, for symmetric
tops, depend in a complex way on $J$, $K$, and $M$~\cite{Haertelt:JCP128:224313}, their effect on
the degree of alignment is highly reduced. According to the Hellmann--Feynman
theorem~\cite{Hellmann:EinführungInDieQuantenchemie:1937, Feynman:PhysRev56:340}, the degree of
alignment is the derivative of the full field-dependent energy with respect to the interaction
parameter. Non-harmonic energy terms are independent of this parameter and thus do not contribute to
the derivative. However, the well-depth terms -- constant offsets in the harmonic-librator energy --
do contribute to the derivative and must be included when differentiating the full energy.
Consequently, the degree of alignment is governed primarily by the harmonic oscillator terms,
supplemented by these well-depth contributions.

\begin{figure}
   \includegraphics[width=\linewidth]{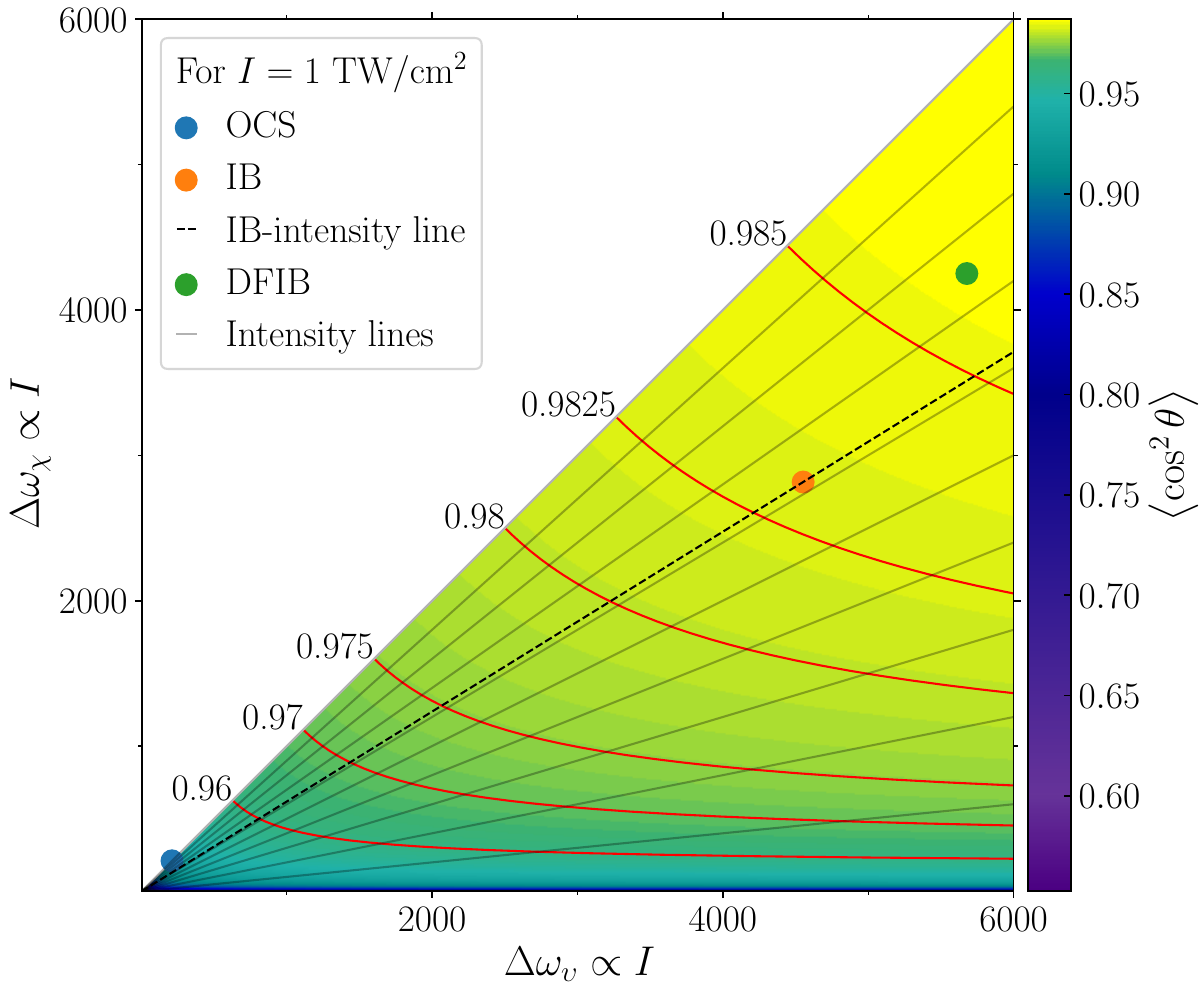}
   \caption{Degree of alignment as a function of $\Delta\omega_\upsilon$ and $\Delta\omega_\chi$ for
      $b=1$ and $\Trot=0$~K. Circles: \textcosthreeD for specific molecules evaluated at
      $\Ialign=1$~\TWpcmcm. Red lines: constant-alignment lines with the corresponding degree of
      alignment indicated next to the diagonal. Grey lines: \textcosthreeD as a function of
      intensity. Black dashed line: \textcosthreeD as a function of intensity for \IB.}
   \label{fig:2D-map}
\end{figure}

Molecules with optimal parameters for a given experiment can easily be identified based on the
derived analytical expressions. In the context of 3D alignment, the behavior of all molecules
exhibiting the aforementioned symmetry properties can be characterized by just four fundamental
parameters given by $\Delta\omega_\upsilon$ , $\Delta\omega_\chi$, $b$, and $\Trot$.
\autoref{fig:2D-map} shows an example of the degree of alignment for the two-dimensional
hypersurface as a function of $\Delta\omega_\upsilon$, and $\Delta\omega_\chi$ for $b=1$ and
$\Trot=0$~K.
Red lines show constant-alignment lines in the degree of alignment with the specific value denoted
next to the diagonal. The degree of alignment of OCS, \IB, and \DFIB, determined in the
harmonic-librator limit, at an intensity of $\Ialign=1$~\TWpcmcm are depicted as circles. \DFIB
exhibits the best alignment characteristics among the studied species at $\Trot=0$~K. OCS, however,
is also a molecule often studied in connection with molecular alignment and
orientation~\cite{Holmegaard:NatPhys6:428, Nielsen:PRL108:193001, Trippel:PRA89:051401R,
  Trippel:PRL114:103003, Kraus:PRL109:233903, DiFraia:PCCP30:19733, Trabattoni:NatComm11:2546,
  Karamatskos:NatComm10:3364, Fleischer:PRL107:163603}, has relatively poor properties. Since both
$\Delta\omega_\upsilon$ and $\Delta\omega_\chi$ are linearly dependent on the intensity, the degrees
of alignment of all molecules as a function of intensity are along straight lines through the origin
and a point evaluated at a specific intensity for a given molecule. For \IB, this line is shown as a
dashed black line. The degrees of alignment of linear molecules, such as OCS, and symmetric-top
molecules are all on the diagonal as $\Delta\omega_\upsilon = \Delta\omega_\chi$. For $b\ne0$ and
finite temperatures, corresponding hyper surfaces can be calculated similarly based on
\eqref{eq:omega_zeta}\ldots\eqref{eq:omega_chi}.

\begin{figure}
   \includegraphics[width=\linewidth]{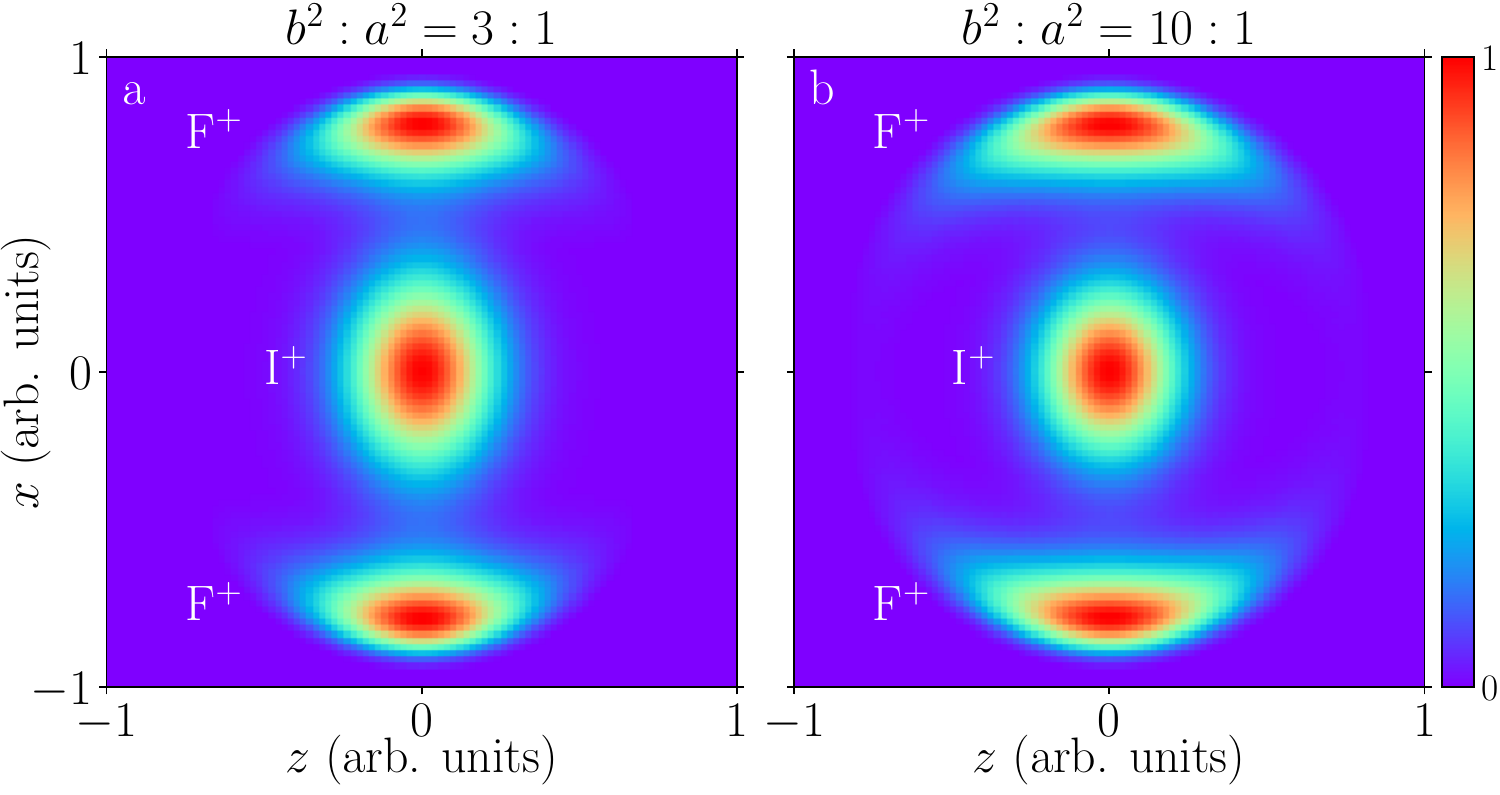}
   \caption{Simulated velocity map imaging detector patterns for \DFIB{} at $\Trot=0.5$~K and
     $\Ialign=1$~\TWpcmcm. Panels (a) and (b) correspond to ellipticities of $b^2:a^2=3:1$ and
     $b^2:a^2=10:1$, respectively. The central feature originates from iodine cations (I$^+$), while
     the peripheral features arise from fluorine cations (F$^+$). The color scale represents
     normalized density.}
   \label{fig:2D-detector-images}
\end{figure}

Very good agreement can also be found in comparison with experimental detector
images~\cite{Nevo:PCCP11:9912} used to characterize 3D molecular adiabatic
alignment. Figure~\ref{fig:2D-detector-images} shows simulated detector images, obtained for
instance by velocity map imaging~\cite{Eppink:RSI68:3477}, for ellipticities of $b^2:a^2=3:1$ and
$b^2:a^2=10:1$ at a rotational temperature $\Trot=0.5$~K for \DFIB.
Both atoms, iodine and fluorine, in the form of cations, are projected onto a detector as would be
the case in a typical velocity-map-imaging spectrometer. The core feature of the structure is
defined by the iodine ions, while the two peripheral features originate from the fluorine ions. The
radial speed for the fragments was 0.9~arb.~units with a variance of 0.001~arb.~units. The peak
intensity is $\Ialign=1$~\TWpcmcm. A multivariate Gaussian structure can be seen in the center for
the iodine ions. The degree of alignment along the $x$-axis is smaller compared to the degree of
alignment along the $z$-axis. This can be attributed to the ellipticity of the laser. For the two
ellipticities, iodine is more confined for $b^2:a^2=10:1$ whereas fluorine is more confined for
$b^2:a^2=3:1$, relative to the other ellipticity. For the fluorine ions, the degree of alignment
along the $z$-axis is worse compared to the structure for the iodine ions. This is due to the
position of the fluorine atoms in the molecule and the fact that the overall rotation about the C--I
axis has the smallest reduced interaction parameter. The degree of alignment of the fluorine atoms
along the $x$-axis is partly given by the libration of the molecule along the angle $\chi$. On the
other hand, the finite velocity distribution comes into play here. Overall, a very good agreement
can be observed in comparison with experimental data~\cite[Fig.~4]{Nevo:PCCP11:9912}.

% Conclusions
We demonstrate that our derived analytical angular distributions accurately characterize the
adiabatic 3D alignment of asymmetric top molecules possessing a polarizability tensor diagonal in
the principal-axes-of-inertia system. The analytical expectation values for the degree of alignment
were compared with numerical results and show excellent agreement within the harmonic-librator limit
for both the low-temperature/small-molecule and high-temperature/large-molecule regimes. The good
agreement for \IB in the high-temperature/large-molecule limit shows that molecules of similar size
behave practically classically already at 0.5~K~\cite{Kumarappan:JCP125:194309}, a rotational
temperature typically representing the lower limit reached in standard supersonic-expansions
molecular beams. Deflector-dispersed beams with even lower effective temperatures may exhibit
deviations from classical behavior, though the magnitude of such deviations remains to be quantified
experimentally. Overall, the derived expressions provide comprehensive insight into 3D alignment as
it relates to molecular properties, laser parameters, and rotational temperature. In addition, the
previously known descriptions for linear and symmetric top molecules in linearly polarized laser
fields can be obtained from the general descriptions presented here using the corresponding limits.

Furthermore, utilizing reduced parameter sets provides a more generalized framework for systematic
analysis. While the current framework assumes a polarizability tensor parallel to the principal axes
of inertia, future work could explore two avenues to broaden applicability: Firstly, relaxing this
constraint using perturbation theory to treat fully asymmetric molecules, like indole, which are
close to the required symmetry but exhibit minor deviations~\cite{Mullins:NatComm13:1431}. Secondly,
developing a precise methodology to correlate pendular-limit states with their field-free
counterparts, similar to those used for asymmetric tops in static fields~\cite{Kanya:PRA70:013403}.

The description of very large molecules at high temperatures should also provide excellent agreement
with classical particle models~\cite{Amin:JACS147:7445} offering highly accurate descriptions also
for such objects in optical tweezers~\cite{Galajda:OptExp11:446, Gao:PNAS116:4018}. Our results are
also ideal for estimating degrees of alignment of molecules using truncated-pulse field-free
alignment using fast shutters~\cite{Yan:JCP111:4113, Underwood:PRL90:223001} or pulse
shaping~\cite{Mullins:NatComm13:1431}. Here, molecules are first aligned adiabatically. A subsequent
quick field turn-off leaves them transiently aligned for a short duration with no electric field
present. Using the sudden approximation, even the field-free dynamics after switching off the pulse
could potentially be captured well.

Overall our findings open up important avenues for planning, understanding, and modeling complex
molecular-frame experiments across a broad range of systems, from small molecules to proteins.

\section{Acknowledgments}
We acknowledge financial support by Deutsches Elektronen-Synchrotron DESY, a member of the Helmholtz
Association (HGF), also for the provision of experimental facilities and for the use of the Maxwell
computational resources operated at DESY. The research was further supported by the Cluster of
Excellence ``Advanced Imaging of Matter'' (EXC~2056, ID~390715994) of the Deutsche
Forschungsgemeinschaft (DFG), the Helmholtz Foundation through funds from the Helmholtz-Lund
International Graduate School (HELIOS, Helmholtz project number HIRS-0018), and from the German
Federal Ministry of Education and Research (BMBF) and the Swedish Research Council (VR~2021-05992)
through the R\"ontgen-\AA ngstr\"om cluster (project 05K22GUA).

\section{Data availability}
As this is a theoretical study, all results are derived from analytical derivations and numerical
simulations. Supporting materials are available from the corresponding author upon reasonable
request.

\section{Code availability}
The code for the data analysis is available from the corresponding author upon reasonable request.

\bibliography{string,cmi}

\onecolumngrid
\listofnotes  % set final in class options to not see this list;-)

\end{document}